\documentclass[12pt,a4paper,twoside]{article}
\usepackage{epsfig}
\usepackage{baltlat1}
\usepackage{wrapfig}
\begin{document}
\ \
\vspace{0.5mm}

\setcounter{page}{1}
\vspace{8mm}

\titlehead{Baltic Astronomy, vol.12, XXX--XXX, 2003.}

\titleb{HOMOGENEITY OF BRIGHT RADIO SOURCES AT 15~GHz ON THE SKY AND IN THE SPACE}

\begin{authorl}
\authorb{T.~G.~Arshakian}{1,}\footnote[3]{On leave from Byurakan Astrophysical Observatory,  
      Byurakan 378433, Armenia},
\authorb{E.~Ros}{1},
\authorb{J.~A.~Zensus}{1},
\authorb{M.~L.~Lister}{2}
\end{authorl}

\begin{addressl}
\addressb{1}{Max-Planck-Institut f\"ur Radioastronomie, Auf dem H\"ugel 69, 53121 Bonn, Germany} \\
\addressb{2}{Department of Physics, Purdue University, 525 Northwestern Ave., W.~Lafayette, IN 47907-2036, U.S.A.} 
\end{addressl}

\submitb{Received: 12 October 2003}

\begin{abstract}
A revised sample of the 2~cm Very Long Baseline Array (VLBA) survey is studied to test the isotropic distribution of radio sources on the sky and their uniform distribution in space. The revised sample is complete to flux-density limits of 1.5\,Jy for positive declinations and 2\,Jy for $0^{\rm o}>\delta>-20^{\rm o}$. At present the active galactic nuclei sample comprises 122 members. Application of the two-dimensional Kolmogorov-Smirnov (K-S) test shows that there is no significant deviation from the homogenous distribution in the sky, while the $V/V_{\rm max}$ test shows that the space distribution of active nuclei is not uniform at high confidence level (99.9~\%). This is indicative of a strong luminosity and/or density evolution implying that active nuclei (or jet activity phenomena) were more populous at high redshifts,  $z\sim2$. 
\end{abstract}





\sectionb{1}{INTRODUCTION AND THE SAMPLE}
The original 2~cm sample (Kellermann et al. 1998 and Zensus et al. 2002) was selected from the Stickel catalog (Stickel et al. 1994) with the aim of including flat-spectrum, core-dominated active galactic nuclei with spectral indices $> -0.5$. This sample was enlarged by adding some sources of special interest, such as steep spectrum sources,  a few lobe-dominated sources and gigahertz-peaked spectrum sources. The 2~cm sample is being used to study the morphology of compact radio sources,  their variability and jet kinematics. To investigate statistically significant correlations between the jet parameters, to compare theoretical predictions of relativistic beaming model, it has been necessary to define a complete flux-density limited sample. The revised sample consists of 122 AGN (see e.g., Zensus et al. 2003). The main selection criterion is the flux-density limit at 15~GHz; all variable sources with galactic latitude $|b| > 2.5^{\circ}$ and with measured VLBA flux densities exceeding 1.5 Jy (2 Jy for southern sources) at any epoch since 1994 are included in the sample. All 122 AGN are radio-loud and core-dominated. Most of them have superluminal radio jets on parsec-scales. At present there are 99 members and 23 candidates. Among members there are 73 quasars,  14 BL Lacs, 7 radio galaxies and 5 sources with no optical counterparts. All AGN except 0642+449 ($z=3.41$) have redshifts less than 2.43.

Since this sample is a subsample selected from the Stickel catalog it is important to understand what selection effects it might contain. Here, we investigate the homogeneity of the 2~cm flux-density limited sample on the sky and in the space.

\sectionb{3}{SKY DISTRIBUTION}
If the sample of radio sources is free from selection effects, then we should expect that radio sources are distributed uniformly in the sky.
The distribution of the 99 member AGN on the celestial sphere is shown in Figure 1. 

\vskip1mm
\begin{wrapfigure}{r}[0pt]{80mm}
\centerline{\psfig{figure=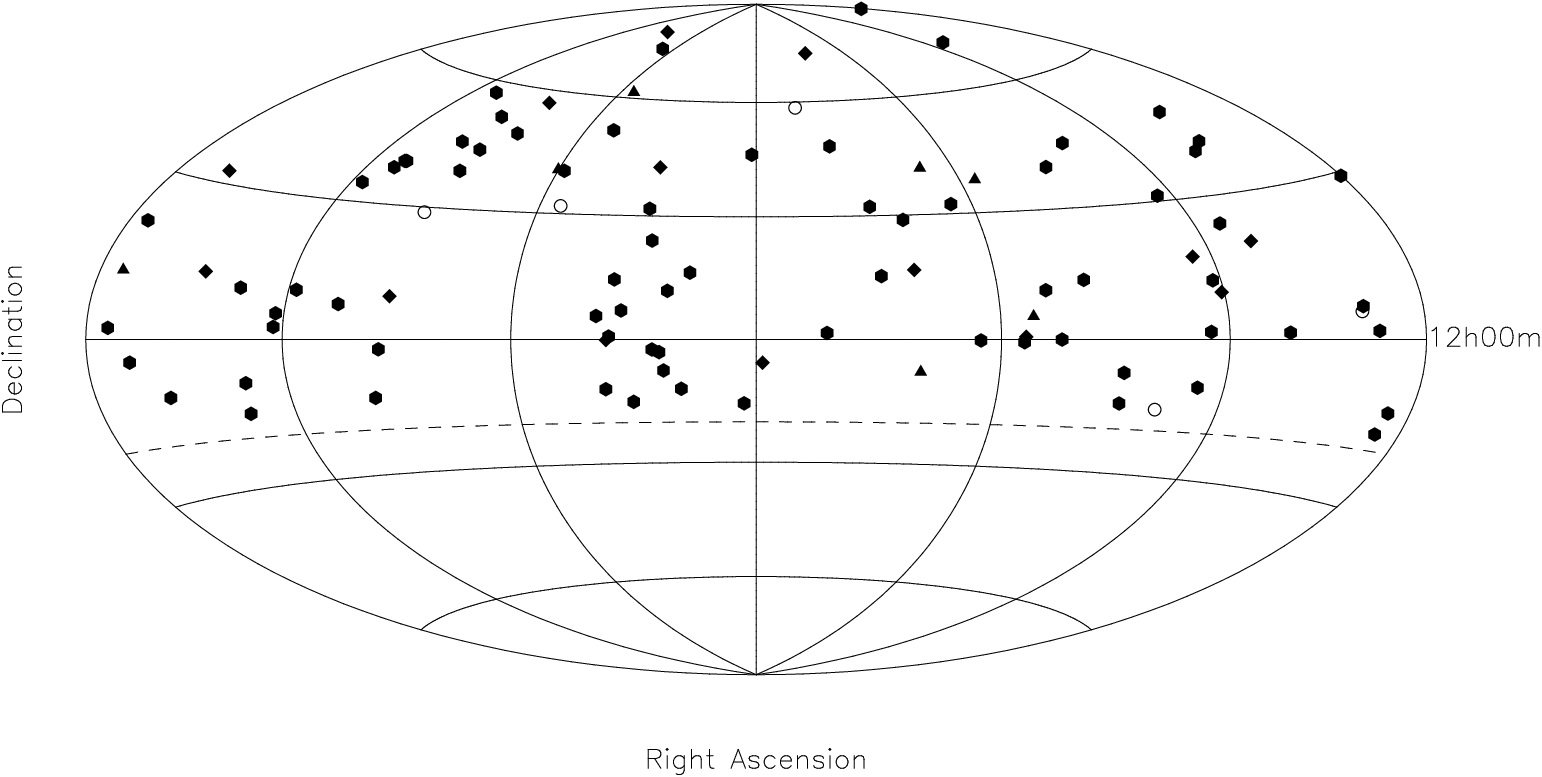,width=41truemm,angle=-90,clip=}}
\captionb{1}{The filled circles, diamonds and triangles are quasars, BL Lacs and galaxies, and open circles are sources with no optical counterparts.}

\end{wrapfigure}

The distribution of radio sources looks patchy with a few clusters and voids in the sky. We also generated 99 points randomly distributed in the same region of the sky and saw that for small number of random points there are modest clusterings and voids as seen in the real data. To confirm this result we performed two dimensional K-S test to test the consistency of the coordinates of sources  with the coordinates of  uniformly distributed points on the sphere. No significant deviation between data and model is found for any redshift bin.     

If sources are randomly distributed in the sky, then the right ascensions will be uniformly distributed and declinations will be $\sin(\pi/2-\delta)$ distributed. The one-dimensional K-S test does not show significant deviation of the declinations from the expected distribution. One interesting result comes for the right ascensions of 15 sources between $1.3<z<2$: there is a clustering of 14 sources between $03^{\rm h}<\alpha<15^{\rm h}$, so, the uniform distribution can be rejected at high, 99.92~\%, confidence level. The non-uniform distribution is still significant for 24 sources with $z>1.3$, but since this result is based on small number statistics, it can only be taken as a provisional conclusion.

\sectionb{4}{SPACE DISTRIBUTION}
We use the distribution of $V/V_{\rm max}$ (Schmidt 1969) to test the null hypothesis of a uniform distribution of AGN in the space. Here $V$ is the volume of space enclosed by the distance of the AGN having certain flux density, and $V_{\rm max}$ is the maximum volume of space within which this source could be observed and still be included in the complete sample. If sources have a uniform distribution in space, then $\langle V/V_{\rm max}\rangle=0.5$.  The difference between the expected and observed values of $\langle V/V_{\rm max}\rangle$ allows the significance of a deviation of real distribution from the uniform to be estimated. 

\begin{wrapfigure}{r}[0pt]{60mm}
\centerline{\psfig{figure=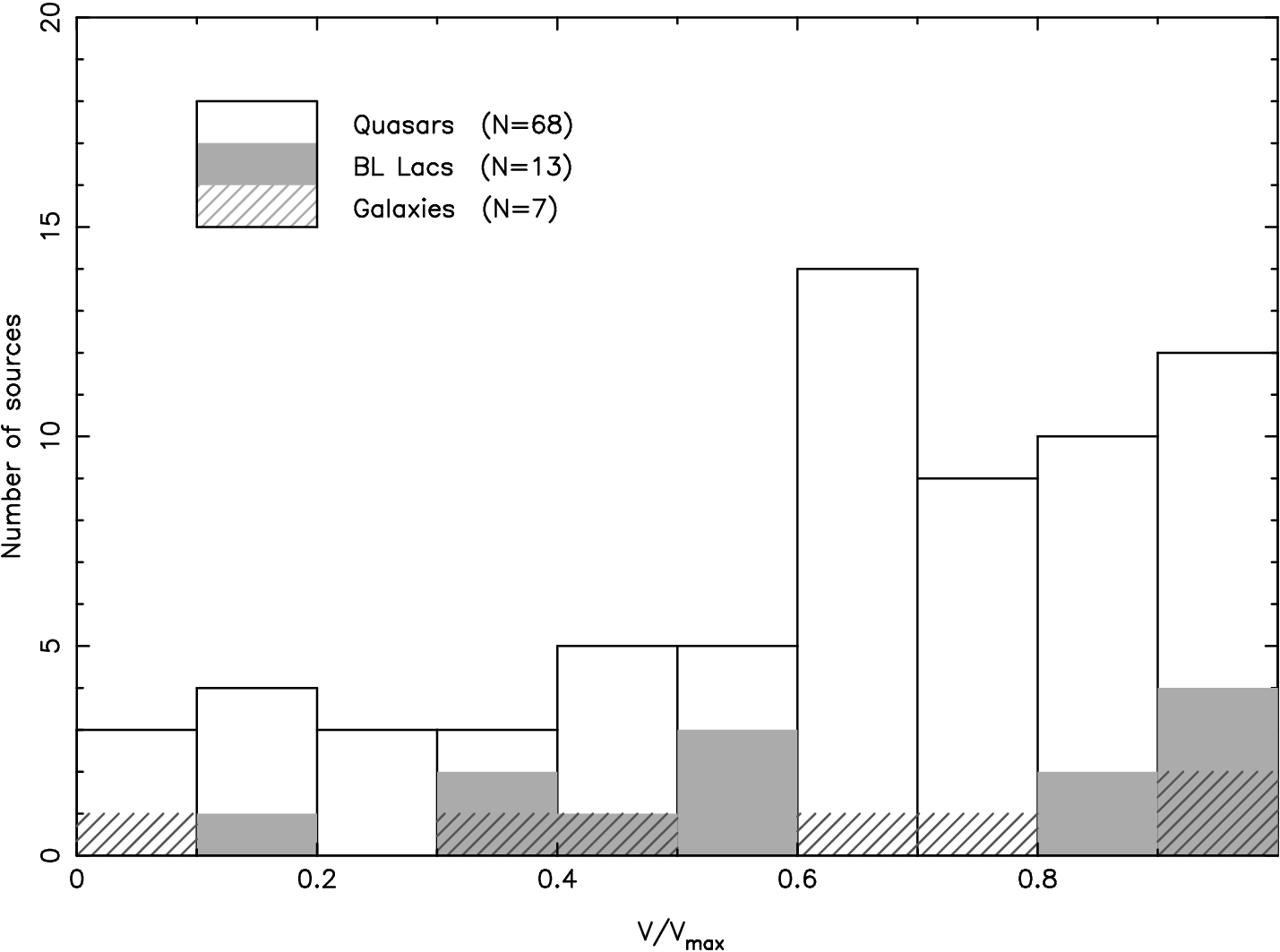,width=40truemm,angle=-90,clip=}}
\captionb{2}{The distribution of $V/V_{\rm max}$ for quasars, BL Lacs and galaxies.}
\end{wrapfigure}

The distribution of $V/V_{\rm max}$ for 88 AGN with redshifts is presented in Figure 2.  For quasars, the distribution is not uniform at high confidence levels (see Table 1), and it is biased towards large $V/V_{\rm max}$ which indicates that quasars, and hence the jet activity phenomena, were more populous at high redshifts, $z\sim2$. The BL Lacs show a similar trend but this is not statistically significant because of the small number of sources. The distribution seems to be uniform for radio galaxies, i.e. with no luminosity/density evolution. This can be understand since all seven radio galaxies occupy the low redshift region where the density/luminosity evolution is negligible, but a larger sample would be needed to confirm this result.
\vskip1mm
\begin{wrapfigure}{r}[0pt]{7.5cm}
\vbox{\footnotesize
\begin{tabular}{lcll}
\multicolumn{4}{c}{\parbox{6cm}{\baselineskip=8pt
~~~~{\smallbf Table 1.}{\small\ $V/V_{\rm max}$ test for 88 AGN ($H_{\rm 0}=70\,\,{\rm km~s^{-1}~Mpc^{-1}}, q_{\rm 0}=0.1$)}}}\\
\tablerule
ID\hhuad &Number &$\langle V/V_{\rm max}\rangle$ & K-S (\%)\\
\tablerule
QSO        &  68 & $0.64\pm0.03$ & 99.9\hhuad\\[-2pt]
BL Lacs    &  13 & $0.64\pm0.07$  & 75\hhuad\\[-2pt]
Galaxies   &   7 & $0.57\pm0.12$  & 16\hhuad\\[-2pt]
\tablerule
\end{tabular}
}
\end{wrapfigure}
The results are in agreement with the value of $\langle V/V_{\rm max}\rangle=0.68\pm0.04$ for steep-spectrum quasars and radio galaxies (Longair \& Scheuer 1970) having extended double structure from the complete 3CRR sample. Dunlop \& Peacock (1990) studied the evolution of the radio luminosity function for extragalactic radio sources combining different flux-density limited, complete samples at 2.7~GHz from the literature. They conclude that both the steep- and flat-spectrum radio sources show the same form of evolutionary behavior: radio luminosity function of quasars and radio galaxies displays positive luminosity evolution out to $z\sim2$, and it declines at high 
redshifts.
 A similar evolutionary behavior up to $z\sim2$ is shown by the bright flat-spectrum radio quasars and BL Lacs from the 2~cm sample. \\


The main conclusions are: (i) the distribution of bright AGN at 15~GHz is uniform on the sky, and (ii) the distribution of quasars and BL Lacs is biased towards large redshifts which is indicative of a strong luminosity and/or density evolution.

\vskip7mm
ACKNOWLEDGMENTS.\ We thank Marshall Cohen for valuable comments, Alan Roy and the members of 2~cm survey team for discussions. TGA is grateful to the AvHF for the award of a Humboldt Post-Doctoral Fellowship.

\goodbreak

\References
\ref
Dunlop~J.~S., Peacock,~J.~A. 1990, MNRAS, 247, 19
\ref
Kellermann~K.~I., Vermuelen~R.~C., Zensus~J.~A., Cohen~M.~H. 1998, AJ, 115, 1295
\ref
Longair~M.~S., Scheuer~P.~A.~G. 1970, MNRAS, 151, 45
\ref
Schmidt~M. 1969, ApJ, 151, 393
\ref
Stickel~M., Meisenheimer~K., Kuehr~H. 1994, A\&AS, 105, 211
\ref
Zensus~J.~A., Ros~E., Kadler~M., et al. 2003, in {\smalit Radio Astronomy at the Fringe}, ASP Conf.\ Ser.\ 300, eds.  J.A.~Zensus, M.H.~Cohen, E.~Ros,\ p.\,27
\ref
Zensus~J.~A., Ros~E., Kellermann~K.~I.,\, et al. 2002, AJ, 124, 662

\end{document}